\documentstyle[aps,epsf,rotate]{revtex}
\font\mb=msbm10

\begin{document}

\draft
\title{Fractal fractal dimensions of deterministic transport coefficients}
\author{R.Klages$^1$ and T.Klau{\ss}$^{1,2}$}
\address{$^1$ Max-Planck-Institut f\"ur Physik komplexer Systeme,
N\"othnitzer Str. 38,  D-01187 Dresden, Germany\\
$^2$ TU Dresden, Fachrichtung Mathematik, D-01062 Dresden, Germany\\
E-mail: rklages@mpipks-dresden.mpg.de and klausz@mpipks-dresden.mpg.de}
\date{\today}
\maketitle
\begin{abstract}
If a point particle moves chaotically through a periodic array of scatterers
the associated transport coefficients are typically irregular functions under
variation of control parameters. For a piecewise linear two-parameter map we
analyze the structure of the associated irregular diffusion coefficient and
current by numerically computing dimensions from box-counting and from the
autocorrelation function of these graphs. We find that both dimensions are
fractal for large parameter intervals and that both quantities are themselves
fractal functions if computed locally on a uniform grid of small but finite
subintervals. We furthermore show that there is a simple functional
relationship between the structure of fractal fractal dimensions and the
difference quotient defined on these subintervals.
\end{abstract}

\section{Introduction}
Endeavours to understand fundamental laws of nonequilibrium statistical
mechanics starting from the deterministic, chaotic equations of motion of a
many-particle system led to the discovery of specific fractal structures
forming the link between microscopic and macroscopic scales: For dissipative,
so-called thermostatted systems in which deterministic transport processes
such as heat or shear flow, or electric conduction, are generated by external
fields, fractal attractors were claimed to be at the origin of the second law
of thermodynamics \cite{HHP87,EvMo90,HoB99}. For open Hamiltonian systems the
associated repeller usually exhibits fractal properties
\cite{GaDo95,Gasp,Do99}, and in case of closed Hamiltonian systems the
hydrodynamic modes of diffusion and reaction-diffusion were found to be
fractal \cite{GDG01,GCGD01,ClGa02}. In all three cases the fractal dimensions
of these irregular structures could be explicitly linked to the transport
coefficients of the different systems \cite{GDG01}.

Singular measures and their fractal structures in phase space were also
discovered to play a fundamental role for the entropy production in simple
two-dimensional area-preserving multibaker maps \cite{Gasp,Do99,TG2}.
Furthermore, for a particle moving chaotically through a periodic array of
scatterers the associated transport coefficients of drift, diffusion and
chemical reaction were found to be irregular, typically fractal functions of
control parameters
\cite{MH87,LNRM95,RKD,RKdiss,KlDo99,GaKl,HaGa01,HaKlGa02,KoKl02}. However, in
the latter case a detailed assessment of the irregularity of these curves in
terms of fractal dimensions was not yet performed. One reason for the lack of
such an analysis was the limited size of the corresponding data sets due to
the fact that the precise computation of such irregular transport coefficients
is generally very tedious. More recently, an exact solution became available
for the irregular transport coefficients of a two-parameter piecewise linear
chaotic map defined on the unit interval and periodically continued on the
line \cite{GrKl02}, see Section II. In this case the explicit expressions for
the diffusion coefficient and for the current were obtained in form of coupled
recursion relations that converge very quickly enabling to numerically
generate, in principle, arbitrarily precise and large data sets.

By using this algorithm, in this paper we analyze the structure of the
parameter-dependent diffusion coefficient and of the current by numerically
computing the box-counting dimension as well as the dimension related to the
autocorrelation function of these graphs \cite{Falc90,Tric95}. These two
methods and our numerical implementation of them are briefly described in
Section III. As is well-known \cite{RKD,RKdiss,KlDo99}, the type of
irregularity of these curves changes by changing the parameter: For example,
in certain parameter regions the diffusion coefficient shares some features
with the self-similar Koch curve, whereas in other regions it looks like a
deformed Takagi function \cite{Tak03} or resembles some Weierstra{\ss}
function \cite{BeLe80}. This appearance of different seemingly fractal
structures in different subintervals motivates us to compute dimensions not
only {\em globally} for large intervals, but also {\em locally} by dividing
these large intervals uniformly into a grid of small but finite
subintervals. This enables us to study the dimensions as functions of the
position of these subintervals. Our main result is presented in the two first
parts of Section IV and states that, firstly, both transport coefficients
possess fractal dimensions on large scales, and secondly, that both dimensions
are fractal functions with respect to the position of the respective
subintervals on which they are computed. Hence, we say that these transport
coefficients are characterized by {\em fractal fractal dimensions}.

In the same section we compare the results for these two dimensions with the
one obtained from a third method that amounts in computing the difference
quotient of the fractal diffusion coefficient with respect to the same grid of
small but finite subintervals. Previously, in case of diffusion this quantity
was found to exhibit a certain structure as a function of the respective
control parameter that appeared to be related to the structure of the
diffusion coefficient \cite{RKdiss}. In this paper we apply a refined analysis
along these lines in parallel to our dimension computations. We find that the
local difference quotient exhibits a structure that is qualitatively
strikingly analogous to the functional dependence of the two local
dimensions. That the two dimensions considered here are closely related to
each other, though, at least in practice, quantitatively not necessarily
identical, is well-known \cite{Falc90,Tric95}. However, the qualitative
analogy to the difference quotient as a function of the position of the
subintervals suggests a further relationship to this additional quantity
which, particularly, is much more easy to compute than the two dimensions. In
Section IV.C we provide a heuristic argument yielding a straightforward
functional relation between the autocorrelation dimension and the difference
quotient that we corroborate numerically. In Section IV.D we critically assess
some subtle numerical problems related to our dimension computations,
particularly by comparing our results for the transport coefficients to the
ones obtained from a respective analysis of the Takagi function. We show that
the latter is a fractal that does not exhibit fractal fractal dimensions and
that our functional relation between the autocorrelation dimension and the
difference quotient does not hold in this case. One may suspect that in this
respect the fractal transport coefficients under discussion are different from
self-affine fractal functions of Takagi \cite{Tak03}, de Rham \cite{dRh57}, or
Weierstra{\ss} \cite{BeLe80} type. Section V contains a summary of our results
and lists some interesting open questions.

\section{A simple map with fractal transport coefficients}
Simple models exhibiting deterministic diffusion are one-dimensional maps
defined by the equation of motion
\begin{equation}
x_{n+1}=M_{a,b}(x_n) \quad , \label{eq:eom}
\end{equation}
where $a,b\in \hbox{\mb R}$ are control parameters and $x_n$ is the position
of a point particle at discrete time $n$. $M_{a,b}(x)$ is continued
periodically beyond the interval $[-1/2,1/2)$ onto the real line by a lift of
degree one, $M_{a,b}(x+1)=M_{a,b}(x)+1$. The map that was studied in Refs.\
\cite{RKD,RKdiss,KlDo99,GrKl02} is defined by 
\begin{equation}
M_{a,b}(x)=ax+b \quad , \label{eq:mapg}
\end{equation}
where $a$ stands for the slope and $b$ for the bias of the map. A sketch of
this map is shown in Fig.\ \ref{fig1}. The Lyapunov exponent of this system is
given by $\lambda=\ln a$ implying that for $a>1$ the dynamics is chaotic.  Let
$\rho_n(x)$ be the probability density for an ensemble of moving particles
evolving according to the Frobenius-Perron continuity equation \cite{Ott}
\begin{equation}
\rho_{n+1}(x)=\int dy \: \rho_n(y) \:\delta(x-M_{a,b}(y)) \quad . \label{eq:fp} 
\end{equation}
Here we are interested in the deterministic current and diffusion coefficient
defined by the first and second cumulant
\begin{equation}
J(a,b):=\lim_{n\to\infty}\frac{1}{n}<x> \label{eq:dkdef}
\end{equation}
and
\begin{equation}
D(a,b):=\lim_{n\to\infty}\frac{1}{2n}(<x^2>-<x>^2) \label{eq:cudef}\quad ,
\end{equation}
respectively, where the angular brackets denote an average over the
probability density $\rho_n(x)$ as a solution of Eq.\ (\ref{eq:fp}) for the
map Eqs.\ (\ref{eq:eom}),(\ref{eq:mapg}).  The upper panels of all figures
shown in this paper depict the parameter-dependent diffusion coefficient
$D(a,b)$ or the current $J(a,b)$ of this map either as functions of $a$ for
zero or fixed bias, or as functions of $b$ for fixed $a>1$. The upper curves
in Figs.\
\ref{fig1} (a) to \ref{fig3} (c) were first computed in Refs.\
\cite{RKD,RKdiss,KlDo99} by means of a numerical implementation of analytical
(transition matrix) methods. In this paper all data sets were generated by the
extremely efficient algorithm described in Ref.\ \cite{GrKl02} yielding data
sets that are precise up to the limits of computer precision, see this
reference for further details.

In Refs.\ \cite{RKD,RKdiss,KlDo99,GrKl02} evidence was provided that all these
functions exhibit a {\em nontrivial structure on arbitrarily fine scales},
which here we consider as a {\em qualitative} definition for a function to be
{\em fractal} \cite{Falc90,BeLe80,Beck}. Partly such a behaviour was verified
by producing blowups of finer and finer parameter regions constantly
exhibiting irregular structures, partly parameter regimes could be detected
that yielded approximately self-affine \cite{Man82} structures, cp.\ to the
upper panels of Figs.\ \ref{fig1} to \ref{fig3}. The physical origin of this
fractality can be traced back to the existence of long-time dynamical
correlations in the deterministic dynamics of Eqs.\
(\ref{eq:eom}),(\ref{eq:mapg}). A particular property of this model is that it
is topologically unstable under parameter variation. That is, depending on the
specific choice of parameters certain orbits may be allowed or forbidden,
which in the theory of symbolic dynamics is known as ``pruning'' of orbits,
respectively of the associated symbol sequences \cite{CAMTV01}. In Refs.\
\cite{RKD,RKdiss,KlDo99} for a certain parameter interval specific series of
points could be identified at which the dynamics is drastically changed
related to this property. These values indeed identified approximately similar
regions appearing on different scales. In case of diffusion, another approach
to understand these fractal structures employs a Green-Kubo formula
\cite{HaKlGa02,KoKl02,KlKo02}. By systematically evaluating the velocity
autocorrelation function of the model the diffusion coefficient can be written
as a series whose single terms represent dynamical correlations of
increasingly higher order. The convergence rate of this series turns out to be
parameter dependent hence assessing the irregular structure of these
curves. For the map under consideration, a limiting case of this approach
enables to analytically relate the shape of the diffusion coefficient to de
Rham-type fractal functions of which the famous Takagi function is a special
case \cite{RKdiss,GaKl}. However, these transport coefficients appear to be
especially complicated fractals in the sense that in different parameter
regions different types of fractal structures show up, cp.\ the upper panels
of all figures. Even more, these structures do not seem to obey simple scaling
laws, in contrast to Takagi, de Rham, or Weierstra{\ss} functions
\cite{Tak03,dRh57,BeLe80}, which is related to the fact that their shapes are
getting deformed as a function of the parameter. For the diffusion coefficient
this particular property may physically be understood with respect to the
Green-Kubo formula mentioned above showing that this transport coefficient
emerges from two different fractal structures that are coupled to each other
via integration \cite{RKdiss,GaKl,KoKl02,KlKo02}. For the current there is
only one of these two sources of fractality, however, even this one changes in
a very intricate way as a function of the parameter \cite{RKmapg} and does not
appear to obey a de Rham-type functional equation
\cite{GrKl02,dRh57}.

Thus, for the map Eqs.\ (\ref{eq:eom}),(\ref{eq:mapg}) there is already quite
some evidence for a non self-affine fractality of the associated transport
coefficients according to the qualitative definition mentioned above. However,
the irregularity of a curve can also be determined by computing dimensions
such as the box-counting dimension. With respect to such quantities, {\em a
set may be called fractal if a dimension can be assigned to it which is not an
integer} yielding a {\em quantitative} definition of this property
\cite{Falc90,BeLe80,Ott,Man82}. In this paper we focus on assessing the
irregularity of the transport coefficients of the map Eqs.\
(\ref{eq:eom}),(\ref{eq:mapg}) particularly with respect to this second
definition. Note that we consider both definitions only as being operational;
concerning discussions about a precise definition of {\em fractal} we refer to
Refs.\ \cite{Falc90,Man82}.

A previous computation of the box-counting dimension for the function
presented in Fig.\ \ref{fig1} (a) based on a data set of $8000$ points led to
the preliminary result that this dimension should be larger than, but very
close to one \cite{RKdiss,KlDo99}. This implies that any dimension
computations are extremely delicate in order to provide evidence for a
possible non-integer dimension. In the following section we briefly describe
two standard methods for computing the dimension of functions as well as our
approach by means of the difference quotient before in the Section IV we
confirm the above statement about a non-integer dimension by applying these
methods.

\section{Methods for assessing the structure of irregular functions}
\subsection{Box-counting dimension}
Let $N(\epsilon)$ be the number of squares needed to cover the graph
${G}\subset\hbox{\mb R}\times\hbox{\mb R}$ of a function $f:\hbox{\mb
R}\rightarrow\hbox{\mb R}$, where $\epsilon$ is the length of one side of a
square.  If $N(\epsilon)$ behaves like a power law for small enough
$\epsilon$, the box-counting dimension $B$ is defined as
\begin{equation}
B:=\lim_{\epsilon\to\infty}-\frac{\ln N(\epsilon)}{\ln \epsilon}\quad
. \label{eq:bc}
\end{equation}
Hence, in order to compute the box-counting dimension of a function $f$ for a
given interval $[c_1,c_2]$ one must count the number of boxes hit by the graph
$G\subset([c_1,c_2]\times f([c_1,c_2]))$ for a quadratic grid of grid size
$\epsilon$. By successively decreasing the grid size, $B$ is obtained from the
slope of a linear regression of Eq.\ (\ref{eq:bc}). Since numerically a graph
always consists of a finite number of points a double-logarithmic plot of
$N(\epsilon)$ as a function of $\epsilon$ may show nonlinear behaviour in the
region of small and large $\epsilon$. To find the linear portion of the
resulting function is one of the crucial problems in order to minimize the
computational error for the box-counting dimension; for further problems see,
e.g., Refs.\ \cite{Falc90,Tric95,Ott,PRS}.

We have numerically implemented this method by generating from a given graph
$G$ for any $\epsilon$ a vector in which each element consists of a pair of
integers referring to the coordinates of the respective box in the grid of
squares that contains this point. This vector was sorted first in $x$, and if
the $x$ coordinates were identical also in $y$. Then it was scanned again, and
if neighbouring pairs of unequal coordinates were encountered an integer
counter was increased by one. 

\subsection{Dimension from the autocorrelation function of a graph}
Let $f:\hbox{\mb R}\rightarrow\hbox{\mb R}$ be a continuous bounded function. If, as in
our case, $f$ is only given on a finite interval $[c_1,c_2]$, this function is
thought to be periodically extended on $\hbox{\mb R}$. The autocorrelation function
of $f$ is then defined by \cite{Falc90}
\begin{equation} 
K(h)=\lim_{c\to\infty}\frac{1}{2c}\int\limits_{-c}^c dx\left(f(x+h)-\bar{f})(f(x)-\bar{f}\right)
\end{equation}
with $\bar{f}:=\lim_{c\to\infty}1/(2c)\int_{-c}^c dx f(x)$ for the average
value of $f$. Eq.\ (\ref{eq:acf}) can be rewritten into
\begin{equation}
K(h)=K(0)-\frac{1}{2}\lim_{c\to\infty}\frac{1}{2c}\int\limits_{-c}^c dx
(f(x+h)-f(x))^2  \label{eq:acf}
\end{equation}
with $K(0):=\bar{f^2}-\bar{f}^2$. If the autocorrelation function satisfies
\begin{equation}
K(0)-K(h)\simeq C h^{4-2A} \label{eq:pld}
\end{equation}   
for small enough $h$ ``it is not unreasonable to expect'' the box-counting
dimension $B$ to equal $A$ \cite{Falc90}. However, it is well-known that this
is not always the case \cite{Tric95}, hence we write the separate symbol $A$
and denote it as the {\em autocorrelation dimension}.\footnote{Note that $A$
is related to the {\em H\"older exponent} by $H=2-A$ \cite{Falc90,Tric95},
which in the context of Brownian motion it is also known as the {\em Hurst
exponent} \cite{PRS}. $A$ may not be confused with the correlation dimension
according to Grassberger and Procaccia \cite{Ott}, which assesses the
fractality of probability measures rather than the one of graphs of
functions.}  As in case of the box-counting dimension, the term $4-2A$ can be
obtained by computing $\Delta K:=K(0)-K(h)$ for successively decreasing $h$
and extracting the slope from a linear regression of this function in a
double-logarithmic plot. Again, the problematic part is to identify the region
of linear behaviour in order to obtain the slope of it.

Note that, in general, the box-counting dimension yields only an upper bound
for the Hausdorff dimension of a graph \cite{Tric95}, hence based on the two
dimensions introduced above we cannot conclude about the Hausdorff dimension
of the transport coefficients.

\subsection{Difference quotient on a grid of small but finite subintervals}
Let a function $f$ be defined on the interval $[c_1,c_2]\in \hbox{\mb R}$. Let
us consider a grid of $n$ subintervals of size $\Delta d:=(c_2-c_1)/n$
satisfying $[d_i,d_{i+1}]\:,\:d_{i+1}:=d_i+\Delta d\:,\:i\in\hbox{\mb N}$, with
$d_1=c_1$ and $d_n=c_2$. To any subinterval we associate a position
$d_i+\Delta d/2$ on the parameter line. On any of these subintervals we now
define the local difference quotient
\begin{equation}
Q_{\Delta d}(d_i):=\frac{f(d_i+\Delta d)-f(d_i)}{\Delta d} \quad . \label{eq:diq}
\end{equation}
If $f$ is differentiable the limit $f'(d_i)=\lim_{\Delta d\to0}Q_{\Delta
d}(d_i)$ exists. However, since fractal functions are by definition not
differentiable $Q_{\Delta d}(d_i)$ will not converge for $\Delta d\to 0$ in
this case. Nevertheless, for given small but finite $\Delta d>0$ the quantity
$Q_{\Delta d}(d_i)$ is well-defined, and $Q$ as a function of $d_i$ may yield
some information about the irregularity of $f$.\footnote{In Ref.\
\cite{RKdiss} $Q_{\Delta d}(d_i)$ was denoted as a ``pseudo-derivative''.} Since
in the following $\Delta d$ will normally be very small and held fixed we drop
all indices and simplify $Q(d)\equiv Q_{\Delta d}(d_i)$.

\section{Results}
We arrange our numerical results in two parts: In Section IV.A we focus on the
case of zero bias $b=0$ of our model Eqs.\ (\ref{eq:eom}),(\ref{eq:mapg}) for
which there is only a nontrivial diffusion coefficient as a function of the
slope $a$ as a parameter. Here we present results for a large parameter
interval as well as for some magnifications of it. In Section IV.B we study
the case of a non-zero bias $b>0$ for which, additionally, there exists a
current. In this case we deal with the diffusion coefficient as well as with
the current as functions of the two parameters $a$ and $b$. In Section IV.C we
clarify the relation between the dimensions and the difference quotient
analytically and numerically. In Section IV.D we furthermore discuss the
Takagi function and argue that it shows properties that are significantly
different from the ones exhibited by fractal transport coefficients.

\subsection{Diffusion coefficient for zero bias}
\begin{figure}
\begin{center}
\epsfxsize=14cm
\centerline{\hspace*{2cm}\rotate[r]{\epsfbox{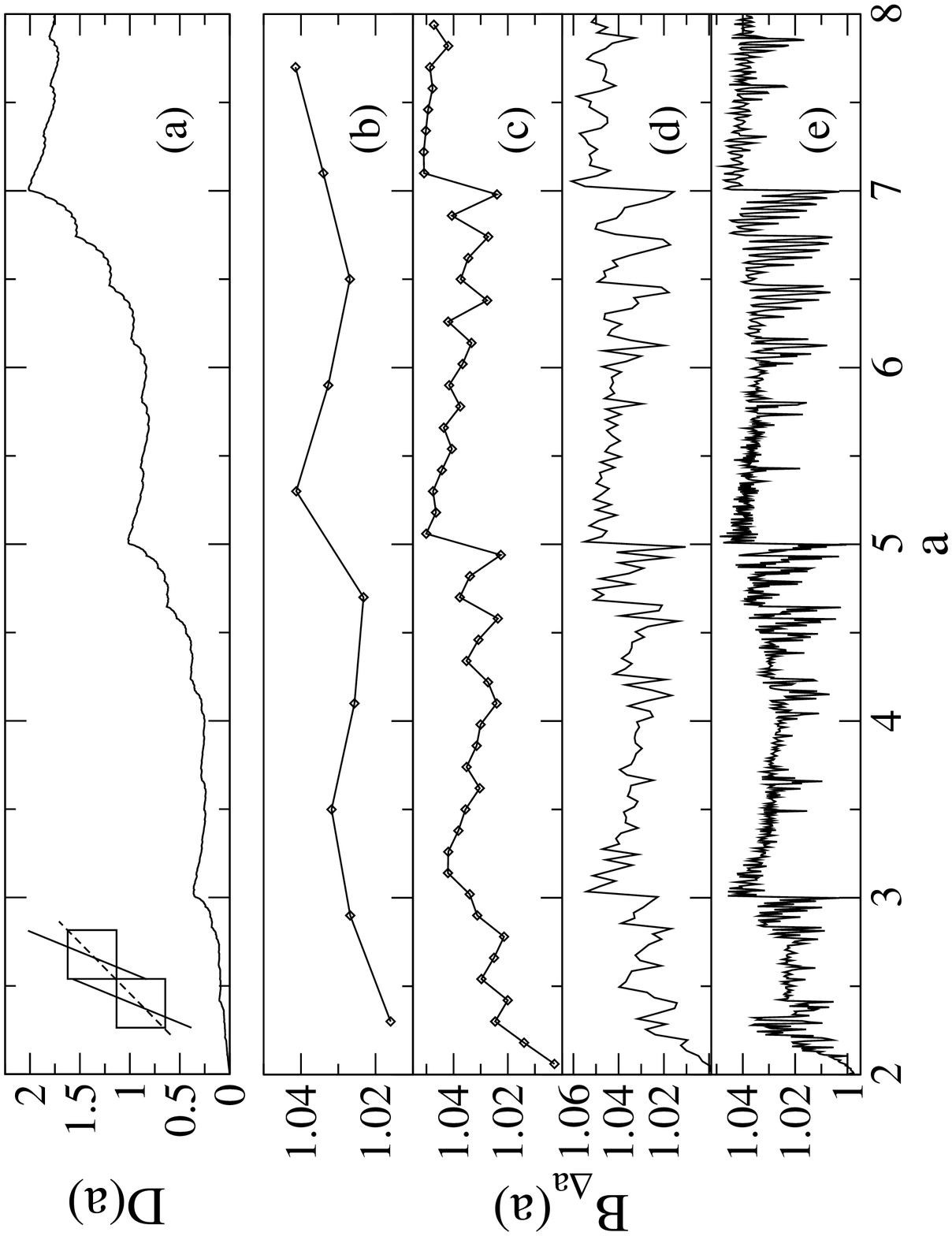}}}
\end{center}
\caption{(a) Diffusion coefficient $D(a)$ on the interval $2\le a \le 8$ with
$2000$ data points for the one-dimensional map Eqs.\
(\ref{eq:eom}),(\ref{eq:mapg}) sketched in the upper left edge with bias $b=0$
and $a$ as the slope of the map. It has a box-counting dimension of $B(D;2\le
a\le 8)=1.039\pm 0.001$ and an autocorrelation dimension of $A(D;2\le
a\le8)=1.074\pm 0.001$. The box-counting dimension was furthermore computed
locally on a regular grid of subintervals of size $\Delta a$. Figs.\ (b) to
(e) depict this local dimension for $\Delta a=0.6,0.12,0.03,0.006$ as a
function of the center $a$ of the subintervals. By decreasing the size of the
subintervals this function is itself getting irregular on finer and finer
scales. Values below one for $a\to 2$ are due to numerical errors.}
\label{fig1}
\end{figure}

We first quantitatively assess the irregularity of the diffusion coefficient
$D(a)$ for the bias $b=0$ on the interval $2\le a\le 8$ as shown in Fig.\
\ref{fig1} (a) by computing values for the box-counting dimension $B$ and the
autocorrelation dimension $A$. In order to get numerically reliable values we
studied the dependence of the two dimensions on the size of data sets with
uniformly distributed points. For $B$ we considered up to $1000000$ data
points of $D(a)$, for $A$ up to $2000000$. We found that the box-counting
dimension converged quickly from below to a value of $B(D;2\le a\le 8)=1.039
\pm 0.001$, where the numerical error denotes the maximum amplitude of the
fluctuations around the asymptotic value for large enough data sets. In
contrast, the autocorrelation dimension $A$ was monotonously decreasing from
above by indicating slow convergence to an asymptotic value. In this case the
data points were fitted by a constant plus power law and the numerical error
was obtained with respect to the upper and lower bounds resulting from
different fit parameters leading to $A(D;2\le a\le 8)=1.074\pm 0.001$. Both
values are obviously not the same, however, both are very close to but
different from one indicating the existence of a fractal according to our
quantitative definition given in Section II. For all the other data sets
presented in the following we were not able to do this tedious error analysis
but computed the dimensions only for a fixed, large number of points. This
implies that, because of the weak convergence of $A$ to an asymptotic value as
mentioned above, results for $A$ are less reliable than for $B$.

We are now interested in the variation of these two dimensions if defined on a
grid of small but finite subintervals. For this purpose we split a given large
parameter interval into a large number of subintervals as described in Section
IV.C. On each subinterval a data set of $1000000$ points was generated in case
of $B$, and of $100000$ points in case of $A$, for which the respective
dimension was computed.  The dimension as a function of the position of these
subintervals we may denote as the {\em local} dimension $B_{\Delta
a}(a)$. First of all, Fig.\ \ref{fig1} (b) to (d) shows that for $D(a)$ on
$2\le a \le 8$ this function is not constant but varies irregularly with
values close to one. This quantifies the observation of Refs.\
\cite{RKD,RKdiss,KlDo99} that $D(a)$ exhibits different types of
fractal structures in different parameter regions.\footnote{In Ref.\
\cite{RKdiss} this property was called {\em multifractal}. However, this
denotation already appears to be occupied for characterizing the irregular
structure of probability measures with respect to exhibiting a spectrum of
non-integer dimensions, see, e.g., Ref.\ \cite{Ott}. Hence, we refrain from
such a denotation.} However, even more Fig.\ \ref{fig1} shows that by
decreasing the size of the subintervals more and more structure in the local
box-counting dimension becomes visible.  This appearance of irregular
structure on finer and finer scales indicates that, according to our first,
qualitative definition of fractality, down to the size of the subintervals the
local box-counting dimension behaves again like a fractal function. How this
function evolves if the size of the subintervals goes to zero may deserve more
detailed investigations that go beyond the scope of this paper.  There exist
lower and upper bounds for the local box-counting dimension, see also Section
IV.D, however, though bounded one may strongly suspect that the limiting case
of $B_{\Delta a}(a)$ for $\Delta a\to 0$ is typically not well-defined. Note
furthermore the somewhat similar oscillatory behavior of $D(a)$ in Fig.\
\ref{fig1} (a) and of the local dimension in Fig.\ \ref{fig1} (e). Analogous
results were obtained for the local autocorrelation dimension.

\begin{figure}
\begin{center}
\epsfxsize=14cm
\centerline{\rotate[r]{\epsfbox{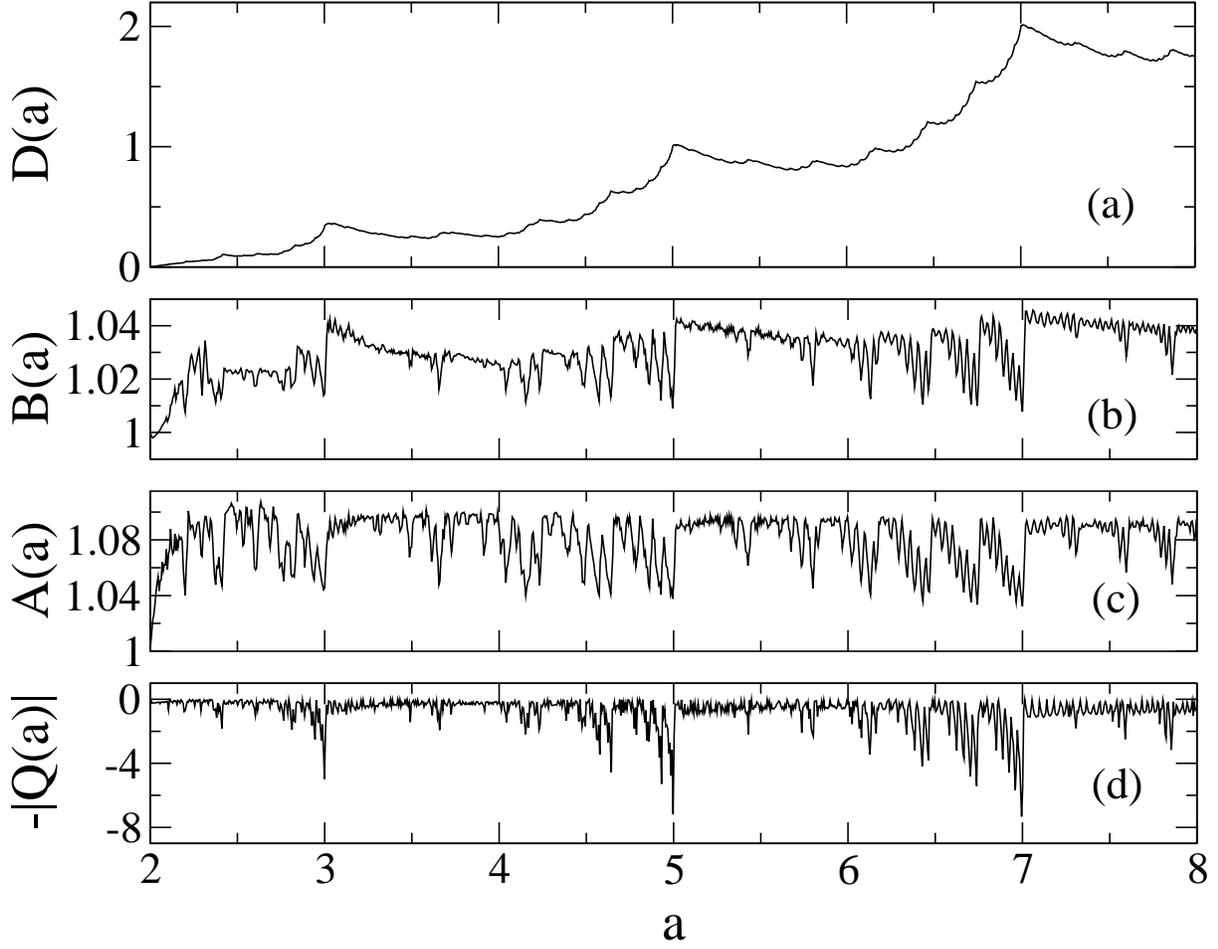}}}
\end{center}
\caption{(a) Diffusion coefficient $D(a)$ as in Fig.\ \ref{fig1}. (b) local
box-counting dimension as in Fig.\ \ref{fig1} (e), however, here as well as
for (c) and (d) a running average was performed over any three neighbouring
points. (c) local autocorrelation dimension $A(a)$, and (d) the negative of
the absolute value of the local difference quotient $Q(a)$. (b) to (d) consist
of $1000$ points each and define the size of the subintervals.}
\label{fig2}
\end{figure}

In Fig.\ \ref{fig2} (b) a slightly ``smoothed'' version of the graph of Fig.\
\ref{fig1} (e) is shown by having performed a running average over any three
neighbouring points. This procedure was applied to most of the following local
quantities in order to eliminate irregularities that may have resulted from
numerical errors. Fig.\ \ref{fig2} (c) and (d) presents respective results for
the local autocorrelation dimension $A(a)$ of $D(a)$ on $2 \le a \le 8$ shown
again in Fig.\ \ref{fig2} (a) as well as for the local difference quotient
$Q(a)$ on the same subintervals according to Eq.\ (\ref{eq:diq}). As for
$Q(a)$, in the following we drop the index $\Delta a$ for $A$ and $B$,
however, the respective values can be obtained from the figure captions. Fig.\
\ref{fig2} (b) to (d) shows a striking similarity between the local
box-counting dimension, the autocorrelation dimension, and the negative of the
absolute value of the local difference quotient. Why we chose particularly the
absolute value of $Q(a)$ will become clear in Section IV.C. The behaviour of
the local difference quotient suggests a somewhat simple functional
relationship between the two local dimensions and this quantity. It seems to
indicate an explanation of the irregular structure of all these curves with
respect to ``differentiating'' the original function of Fig.\ \ref{fig2} (a)
over the set of subintervals, however, as we will discuss in Section IV.D in
general this interpretation has to be handled with much care. Note furthermore
that on finer scales there are some systematic deviations between these three
curves as can be detected, for example, by looking at the envelope of the
extrema of $B(a)$ and $A(a)$ compared with $-|Q(a)|$. Note also the
quantitative deviations between the local box-counting dimension and the local
autocorrelation dimension.\footnote{We remark that, by using the code of Ref.\
\cite{GrKl02}, similar qualitative and quantitative results concerning the
local box-counting dimension and the local difference quotient were obtained
by Z.Koza
\cite{ZiKo02}.}

Due to the limited data sets of $1000$ points for the graphs shown in Fig.\
\ref{fig2} (b) and (c) together with the associated numerical errors we were
not able to compute any reliable values for respective box-counting
dimensions. These problems do not exist for $Q(a)$, however, here the same
difficulty as we discussed for the local dimensions even more clearly shows
up, namely, that this quantity is not well-defined in the limit of small
subintervals. Indeed, by plotting $-|Q(a)|$ for data sets between $1000$ and
$1000000$ points one observes that the local minima keep decreasing thus
indicating that the structure of this curve changes profoundly with respect to
the given number of subintervals. However, by trying to compute the
box-counting dimensions of these different data sets we observed that, over a
significant range of scales, the respective functions obeyed a power law
behaviour, which enabled us to extract values for this dimension. We found
that it monotonously increased from $B(-|Q(a)|;2\le a\le 8)\simeq 1.32$ for
$1000$ points up to $B(-|Q(a)|;2\le a\le 8)\simeq 1.66$ for $1000000$
points. Hence, although $Q(a)$ may not be well-defined in the limiting case of
infinitely small subintervals we conclude that for a finite number of
subintervals and down to certain scales it exhibits characteristics of a fractal
function according to our second, quantitative definition of fractality. Due
to the similarity between the local dimensions and $-|Q(a)|$ one may
conjecture that the same applies if the local dimensions were computed on
finer and finer scales, which appears to be supported by the qualitative
assessment of Fig.\ \ref{fig1}.  Consequently, we say that $D(a)$ as shown in
Fig.\ \ref{fig2} (a) is characterized by a {\em fractal fractal dimension}.

Fig.\ \ref{fig3} presents similar results for $B(a)$, $A(a)$ and $Q(a)$ on
some selected subintervals of $D(a)\:,\:2\le a\le 8$, which exhibit different
fractal structures as shown respectively in the upper panels of (a) to (c). In
(a), though barely visible, there is an underlying triangular-like
self-affinity that reminds a bit of a Koch curve \cite{RKdiss}. The inset of
(b) shows the plot of the Takagi function as computed from the respective
functional equation \cite{Gasp,Do99,Tak03}, which is strictly self-similar
\cite{Man82} on arbitrarily fine scales. (c) may more generally remind of some
Weierstra{\ss} function \cite{BeLe80}. In case (a) the dimension computations
were particularly difficult probably corresponding to the fact that for $a\to
2$ the diffusive dynamics of the model Eqs.\ (\ref{eq:eom}),(\ref{eq:mapg})
approaches a random walk behaviour connected to a smooth functional dependence
of the diffusion coefficient \cite{RKD,RKdiss,KlDo99}. As for Fig.\
\ref{fig2}, we have computed the box-counting dimension
for $D(a)$ on these three subintervals leading to (a) $B(D;2\le a\le
3)=1.021$, (b) $B(D;3\le a\le 3.1)=1.044$ and (c) $B(D;5.6\le a\le
5.7)=1.041$. The numerical error is approximately in the last digits. For the
autocorrelation dimension of (a) to (c) our results indicate that $A>B$. In
all these cases also the two local dimensions as well as the local difference
quotient were computed, see Fig.\ \ref{fig3}, demonstrating again a
considerable similarity between the structure of all these three local
quantities.

\subsection{Diffusion coefficient for non-zero bias}
If we choose for the map Eqs.\ (\ref{eq:eom}),(\ref{eq:mapg}) a non-zero bias
$b>0$ the diffusion coefficient becomes a function of two parameters
$D(a,b)$. Based on stochastic theory one may not necessarily expect that a
diffusion coefficient explicitly depends on the bias, however, in case of this
deterministic chaotic model the nontrivial dynamical correlations mentioned in
Section II cause $D$ to also be a function of $b$. Furthermore, for $b>0$
because of symmetry breaking there exists a nonzero current $J(a,b)$ that,
again, is a function of the two parameters. In the upper panels of Figs.\
\ref{fig4} to \ref{fig5} we present results for the current $J(a)$ at fixed $b$ close to
zero as well as for the diffusion coefficient $D(b)$ and the current $J(b)$
for $a$ close to the onset of chaos at $a=1$, where the Lyapunov exponent of
the system is zero. Typically, the current is a fractal function of the two
control parameters, however, as shown in Fig.\ \ref{fig5} (b) certain
parameter intervals display smooth behaviour, which is reminiscent of
phase-locked dynamics in so-called Arnold tongues \cite{GrKl02}. The inset of
Fig.\ \ref{fig5} (b) depicts a blowup of the current divided by the bias
magnifying the irregular structure of the current on fine scales. The Arnold
tongues also show up in the diffusion coefficient presented in Fig.\
\ref{fig5} (a). Another interesting feature is the existence of {\em current
reversals} in Fig.\ \ref{fig4} (a), that is, depending on the slope $a$ the
sign of the current is either positive or negative reminding of ratchet-like
dynamics \cite{GrKl02,RKmapg}.

As before, for Fig.\ \ref{fig4} (a) and \ref{fig5} the box-counting dimension
$B$ was computed for the whole intervals displayed in these figures yielding
$B(J;2\le a\le8;b=0.01)=1.015$ for Fig.\ \ref{fig4} (a), $B(D;a=1.125;0\le
b\le0.5)=1.021$ for Fig.\ \ref{fig5} (a) and $B(J;a=1.125;0\le b\le0.5)=1.030$
for Fig.\ \ref{fig5} (b), with $B<A$ again in all three cases. Hence, also in
the general case of two parameters both the current and the diffusion
coefficient are fractal functions according to our quantitative definition of
fractality.

For the current of Fig.\ \ref{fig4} (a) as previously the two local dimensions
as well as the local difference quotient were computed, see Fig.\
\ref{fig4} (b) to (d). As far as the regions of negative currents are
concerned, they do not seem to be reflected in specific properties of the
corresponding local quantities. Apart from that, we observe the qualitative
similarity between these three different quantities as already discussed
before. This seems to confirm that there is a simple functional relationship
between $B$, $A$, and the negative of the absolute value of $Q$. However,
Fig.\ \ref{fig5} clearly contradicts this statement by showing that, in these
two cases, $B$ is qualitatively similar to the {\em positive} absolute value
of $Q$. These results indicate that the relationship between the two local
dimensions and $Q$ may indeed be more intricate asking for an explanation of
this sign change. Note furthermore the similarity between $D$, $B$ and, to
some extent, $|Q|$, however, compared with the previous figures it appears that,
by matching the local maxima and minima, for $D$ and for $J$ there are more
deviations between these different structures on finer scales than before.

\subsection{The functional relation between the autocorrelation dimension and the
difference quotient} 
The previous figures provided numerical evidence for the existence of a
relationship between the local box-counting dimension, respectively the
autocorrelation dimension and the local difference quotient. As pointed out in
the introduction, the relation between box-counting and the autocorrelation
function is basic \cite{Falc90}, hence it remains to clarify the relation
of these two quantities to the difference quotient only. Figs.\ \ref{fig2} to
\ref{fig4} show that both $B$ and $A$ are apparently similar to $-|Q|$,
whereas Fig.\ \ref{fig5} clearly tells us that these two local dimensions are
similar to $+|Q|$. This puzzle will be solved in the following based on some
simple analytical arguments.

We start from the observation that the autocorrelation function in form of
Eq.\ (\ref{eq:acf}) already contains the difference $\Delta
f(x,h):=f(x+h)-f(x)$ for a function $f(x)$. Here we are interested in the
autocorrelation function for $f$ on a small parameter interval
$[d-\Delta d/2,d+\Delta d/2]$, thus we rewrite Eq.\ (\ref{eq:acf}) in these
variables and combine it with the assumed power law decay of the
autocorrelation function Eq.\ (\ref{eq:pld}) yielding
\begin{equation}
\frac{1}{2\Delta d}\int_{d-\Delta d/2}^{d+\Delta d/2} dx (\Delta f(x,h))^2=C h^{4-2A}
\quad (h\ll1) \label{eq:ans} 
\quad .
\end{equation}
Let now in addition be $\Delta d\ll1$ and $h\simeq \Delta d$. For small enough
$\Delta d$ we approximate the average over this interval by
\begin{equation}
\frac{1}{\Delta d}\int_{d-\Delta d/2}^{d+\Delta d/2}dx (\Delta f(x,h))^2\simeq
[f(d+\Delta d/2)-f(d-\Delta d/2)]^2\equiv(\Delta f(d))^2 \quad (\Delta d\ll 1)
\end{equation}
leading to
\begin{equation}
\frac{1}{2}(\Delta f(d))^2\simeq C(\Delta d)^{4-2A} \quad . \label{eq:CAQ1}
\end{equation}
However, there is no reason why only $A$ may be a function of $d$, hence
$C\equiv C(d)$, as was already remarked in Ref.\ \cite{Tric95}. By furthermore
noting that $Q(d)\equiv\Delta f(d)/\Delta d$ as introduced in Eq.\
(\ref{eq:diq}), Eq.\ (\ref{eq:CAQ1}) yields the important result
\begin{equation}
|Q(d)|\simeq\sqrt{2C(d)}(\Delta d)^{1-A(d)} \quad . \label{eq:CAQ2}
\end{equation}
This equation predicts a simple relationship between the absolute value of the
difference quotient $Q$ on the left hand side and a combination of $C$ and $A$
on the right hand side as functions of $d$ for fixed subinterval size $\Delta
d$.

We now discuss the validity of this equation for the two cases of (i) $D(a)$
on the interval of $2\le a\le 8$, cp.\ to Figs.\ \ref{fig1}, \ref{fig2}, as
well as for (ii) $D(b)$ on the interval $0\le b\le 0.5$, cp.\ to Fig.\
\ref{fig5} (a). The second panels of Fig.\ \ref{fig6} (a) and (b) from above
show the local autocorrelation dimension $A$ of these two curves, the third
panels the associated local difference quotient $Q$ in its original form,
i.e., without taking the absolute value, whereas the lowest panels contain,
among others, the respective absolute values of $Q$. Eq.\ (\ref{eq:CAQ2})
explains why only the absolute value of $Q$ should be related to $C$,
respectively to $B$. That this is the correct solution is indeed confirmed by
Fig.\ \ref{fig6} by very carefully matching the functions depicted in the
second to the fourth panels of Fig.\ \ref{fig6} (a) to each other, say, just
above $a=7$. Apart from such very fine details, the apparent constant upper
bound of $A(a)$ also points to the absolute value of $Q(a)$.

More important is to understand, on the basis of this equation, this apparent
``sign change'' in the relation between $C$ and $|Q|$ as observed before.  For
this purpose we further simplify Eq.\ (\ref{eq:CAQ2}) in order to check
whether we arrive at a simple linear relationship between these quantities as
suggested by Figs.\ \ref{fig2} to \ref{fig5}. We first expand the right hand
side of Eq.\ (\ref{eq:CAQ2}) according to $(\Delta
d)^{1-A}\simeq1+(1-A)\ln(\Delta d)+\ldots$. The bounds $0\le 1-A\le 0.2$ and
$-10\le \ln(\Delta d)\le -5$ obtained from the numerical data presented in our
figures indicate that we can safely neglect all terms of higher order yielding
\begin{equation}
|Q(d)|\simeq\sqrt{2C(d)}[1+(1-A)\ln(\Delta d)]\quad . \label{eq:lapp}
\end{equation}
For the two cases depicted in Fig.\ \ref{fig6} we checked numerically that
this equation is indeed a very good approximation. Since $\Delta d\ll1$ and
$A\ge1$, the logarithmic term in the above equation is strictly
positive. Hence, the origin of a possible ``sign change'' between $|Q(d)|$ and
$A(d)$ can only be due to the prefactor $C(d)$. The functional dependence of
$C(d)$ indeed turns out to provide the key for a complete understanding: In
the upper panels of Figs.\ \ref{fig6} (a) and (b) $C$ is plotted as a function
of the respective control parameters. In both cases it obviously exhibits an
irregular structure that is quite analogous to the irregular structures shown
by the corresponding quantities $A$, $Q$, and $B$. Hence, $Q$ and $A$ are not
the only functions in Eq.\ (\ref{eq:CAQ2}), respectively Eq.\ (\ref{eq:lapp}),
exhibiting fractal behaviour but there is a third fractal function $C$
involved in the functional relation between them. In a way, this appears to be
natural, since there is no reason why by assuming the power law Eq.\
(\ref{eq:pld}) for the autocorrelation function and extracting a fractal local
autocorrelation dimension from its exponent the respective prefactor should
not also be a fractal. Fig.\ \ref{fig6} shows that in case of the diffusion
coefficient as a function of $a$ for zero bias $b=0$, i.e., if the map Eq.\
(\ref{eq:mapg}) is anti-symmetric, the oscillations of $C(a)$ are somewhat
{\em opposite} to the ones of the corresponding $A(a)$, whereas in case of
symmetry breaking with non-zero bias $b>0$ the oscillations of the respective
function are {\em parallel} to the associated $A(b)$.

Let us first discuss the case where the oscillations of both quantities are
parallel to each other, cf.\ to Fig.\ \ref{fig6} (b). Then there is no
mechanism inverting the oscillations on the right hand side of Eq.\
(\ref{eq:lapp}) that may mimic a minus sign. In case of Fig.\ \ref{fig6} (a)
the situation is more difficult: since the oscillations of $C(a)$ and $A(a)$
are opposite to each other it is not clear in advance which contribution
dominates Eq.\ (\ref{eq:lapp}). However, a quantitative evaluation yields
$1\le A\le 1.13$ and $-10\le \ln(\Delta a)\le -5$, hence
$1\le1+(1-A)\ln(\Delta a)\le 2.3$. On the other hand, with $0\le C\le 40$ it
is $0\le\sqrt{2C}\le 9$; indeed, as shown by Fig.\ \ref{fig6} (a), for large
$a$ intervals it is $C\ge 1$. Hence, one would expect that for most intervals
the first term in Eq.\ (\ref{eq:lapp}) dominates the second one indicating
that the most pronounced features of $|Q(a)|$ are rather related to $C(a)$
than to $A(a)$. A look at the envelope of the largest local maxima of $C(a)$
and $Q(a)$ compared with $A(a)$ confirms this heuristic argument.

In order to assess the validity of our explanation quantitatively, the lowest
panels of Fig.\ \ref{fig6} (a), (b) display the left hand side of Eq.\
(\ref{eq:CAQ2}), that is, the absolute value of the respective local
difference quotient $Q$, in comparison with the right hand side of this
equation $P(d):=\sqrt{2C(d)}(\Delta d)^{1-A(d)}$ before linearization. Both
figures show an excellent agreement of these two functions thus confirming the
validity of Eq.\ (\ref{eq:CAQ2}). Note that in (b) the number of points of
both data sets is not equal indicating that the relation between these two
functions holds over a broad range of subinterval sizes $\Delta b$. The
linearization Eq.\ (\ref{eq:lapp}) is indistinguishable from Eq.\
(\ref{eq:CAQ2}) on the scale of the figures, hence it was not included. We
thus come to the conclusion that in case of the fractal transport coefficients
under discussion indeed the autocorrelation dimension $A$ and the absolute
value of the difference quotient $Q$ are related to each other by a simple,
under certain circumstances approximately linear functional
relationship. However, in addition this relation involves another fractal
function, which is the prefactor of the power law Eq.\ (\ref{eq:pld}) that is
itself a fractal function of the same variable. The oscillations of this
function $C$ can be either parallel or opposite to the oscillations of $A$ and
hence determine whether the oscillations of $|Q|$ are in turn parallel or
opposite to $A$ consequently explaining this sign change that previously was
introduced {\em ad hoc} in the respective relationship.

\subsection{The Takagi function compared with the fractal diffusion coefficient}
An important question is whether the results presented above are specific to
these fractal transport coefficients or whether they more generally apply to
self-affine fractal functions of Takagi, de Rham, or Weierstra{\ss}
type. Below we discuss this point starting from the Takagi function
\cite{Tak03} and then compare our findings with the previous results
providing some insight into the problems with and limitations of assessing a
curve by computing local dimensions and difference quotients.

\begin{figure}
\begin{center}
\epsfxsize=14cm
\centerline{\rotate[r]{\epsfbox{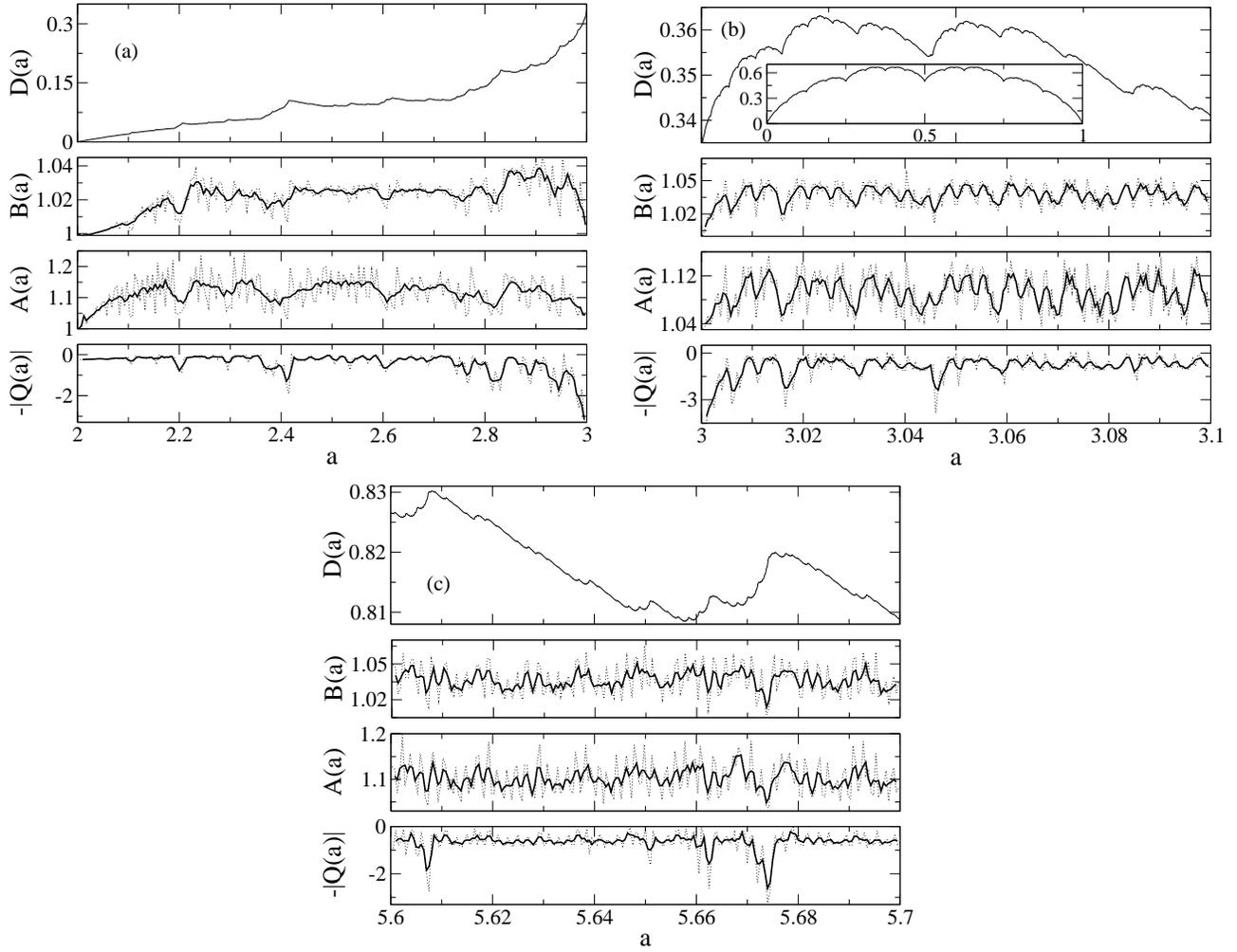}}}
\end{center}
\caption{Diffusion coefficient $D(a)$ for some subintervals of $a$ as well as the
corresponding local box-counting dimension $B(a)$, the autocorrelation
dimension $A(a)$ and the local difference quotient $Q(A)$. In all cases,
$D(a)$ consists of $2000$ points. The inset in (b) depicts, for comparison,
the famous Takagi function. The dotted lines in the lower panels of (a) to (c)
consist of $200$ points each, whereas the solid lines are respectively
smoothed curves obtained from suitable running averages over the original
data.}
\label{fig3}
\end{figure}

\begin{figure}
\begin{center}
\epsfxsize=14cm
\centerline{\rotate[r]{\epsfbox{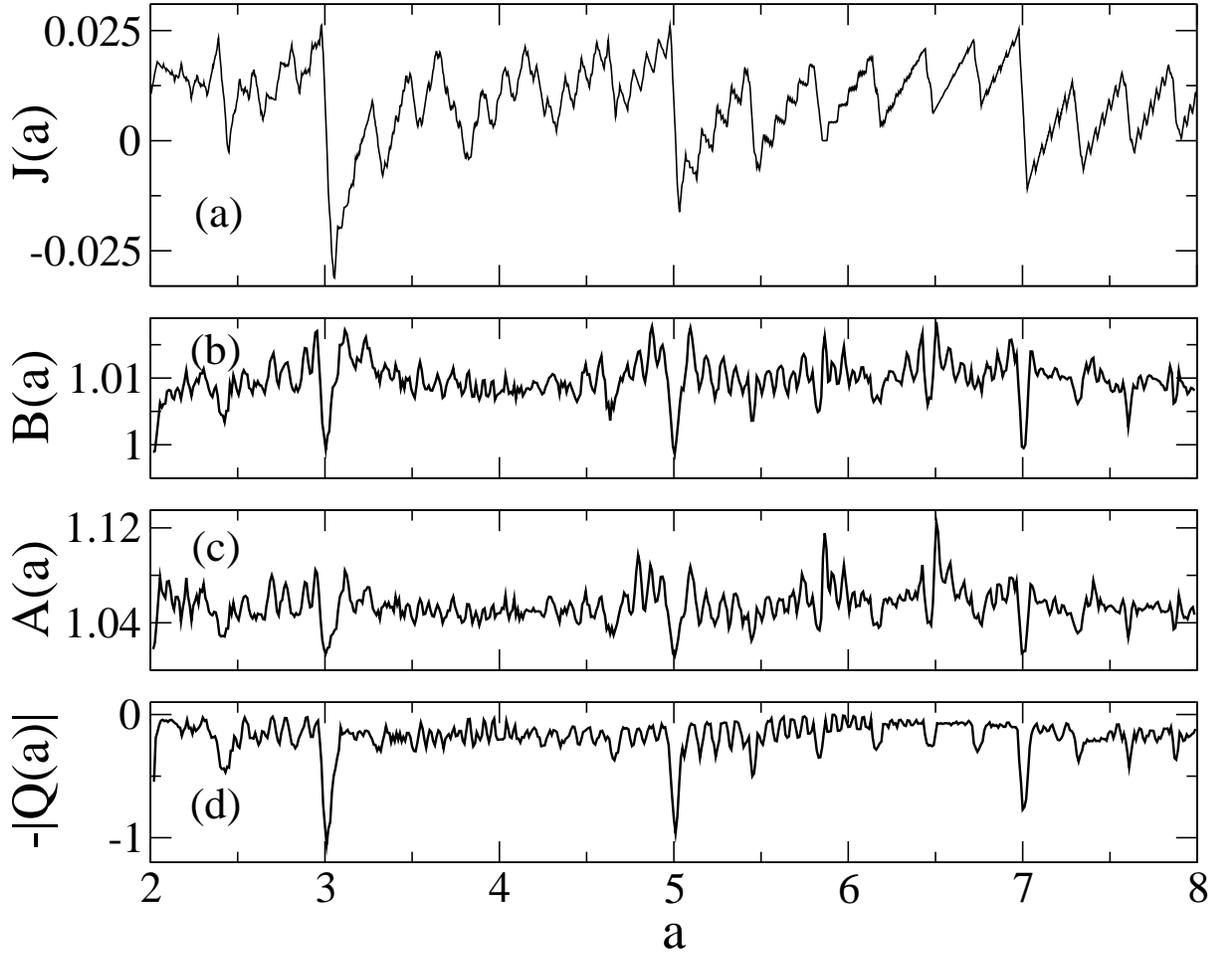}}}
\end{center}
\caption{(a) Current $J(a)$ on the interval $2\le a\le 8$ for $b=0.01$. Note
the regions of negative currents, e.g., above odd integer slopes $a$. As
before, (b) depicts the local box-counting dimension $B(a)$, (c) the local
autocorrelation dimension $A(a)$, and (d) minus the absolute value of the
local difference quotient $Q(a)$. $J(a)$ consists of $2000$ points, the other
three graphs consist of $600$ points each and are smoothed out by running
averages over three neighbouring points.}
\label{fig4}
\end{figure}

\begin{figure}
\begin{center}
\epsfxsize=7.5cm
\centerline{\rotate[r]{\epsfbox{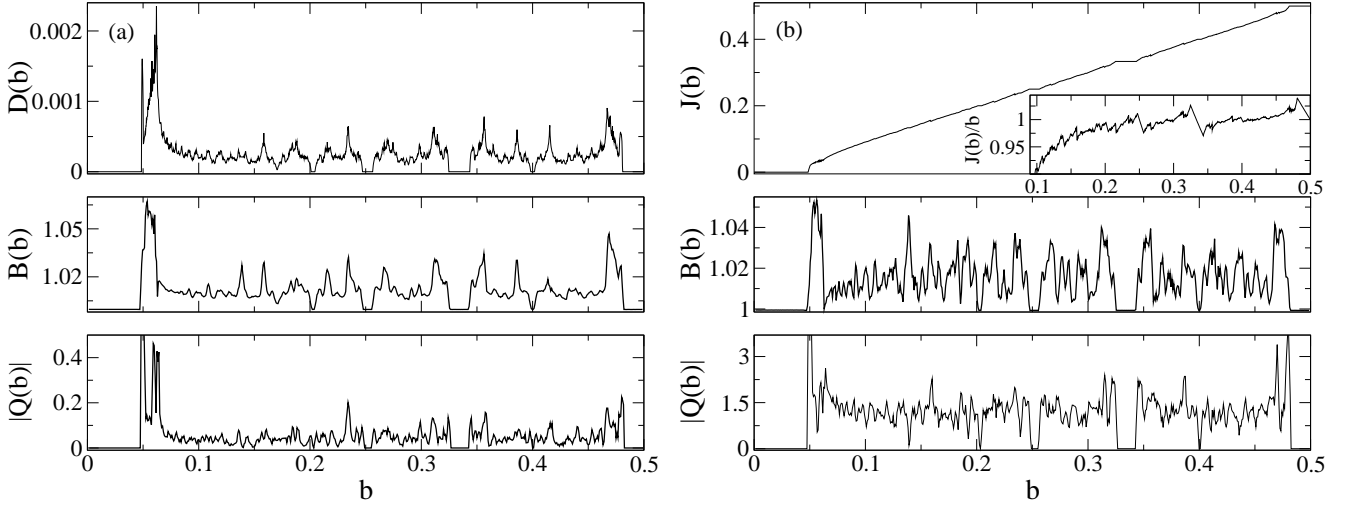}}}
\end{center}
\caption{The upper panel in (a) shows the diffusion coefficient $D(b)$ on the
interval $0\le b\le 0.5$ for $a=1.125$, the respective part in (b) the current
$J(b)$ for the same setting. The inset in (b) presents the current divided by
the bias in order to magnify the fractal fine structure of this function. In
both cases the parameter regions of zero diffusion coefficient, respectively
the plateau regions of the current and the linearly decreasing regions in
$J(b)/b$, correspond to Arnold tongues. The lower two panels in (a) and (b)
depict the local box-counting dimension $B(b)$ as well as, in contrast to the
previous figures, the {\em positive} absolute value of the local difference
quotient $Q(a)$ for both transport coefficients. Both $D(b)$ and $J(b)$
consist of $2000$ points each, the other graphs consist of $600$ points each
and are smoothed like Fig.\ \ref{fig4} (b) to (c).}
\label{fig5}
\end{figure}

\begin{figure}
\begin{center}
\epsfxsize=7.5cm
\centerline{\rotate[r]{\epsfbox{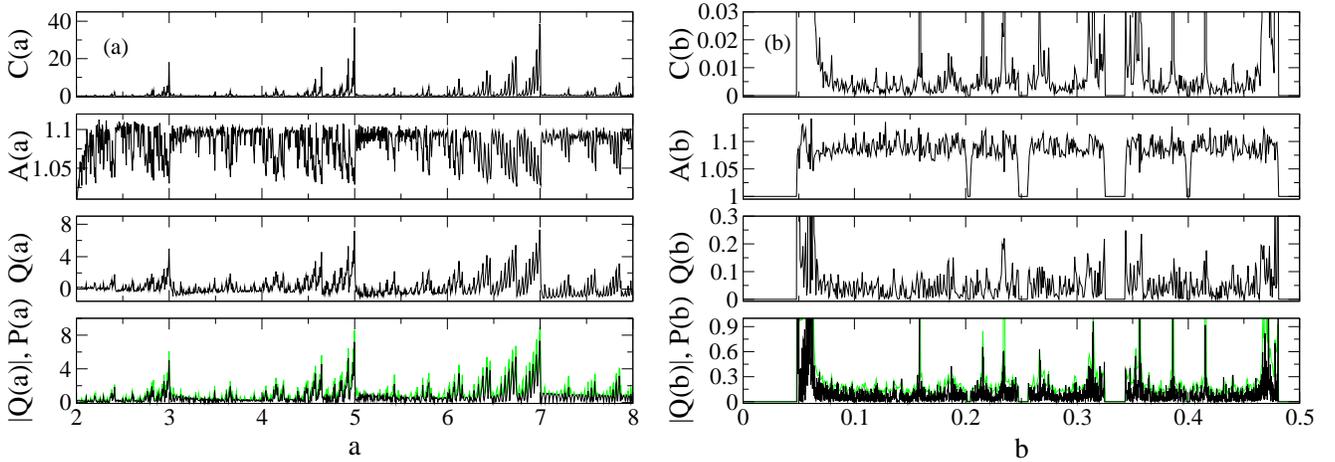}}}
\end{center}
\caption{Assessing the relation between the autocorrelation dimension $A$ and
the difference quotient $Q$ for two representative examples: (a) depicts the
decisive quantities for the diffusion coefficient $D(a)$ on the interval $2\le
a\le 8$, see also Figs.\ \ref{fig1}, \ref{fig2}, whereas (b) shows the same
quantities for the diffusion coefficient $D(b)$ on $0\le b\le 0.5$, cp.\ to
Fig.\ \ref{fig5} (a). The upper panels contain results for the prefactor $C$
of the power law Eq.\ (\ref{eq:pld}), whereas the next panels show again the
local autocorrelation function obtained from the exponent of the same power
law, for (a) cp.\ to Figs.\ \ref{fig1} and \ref{fig2}. Remarkable are the
irregularities of $C$ in both cases and that the oscillations are opposite to
$A$ in (a), whereas they are parallel to $A$ in (b). The third panels depict
again the local difference quotient $Q$, see also Fig.\ \ref{fig2} for (a) and
\ref{fig5} (a) for (b). The fourth panels show a comparison between the
absolute value of $Q$, which is the left hand side of Eq.\ (\ref{eq:CAQ2})
(black line) in comparison to the right hand side $P(a)$ of the same equation
that is a function of $C$ and $A$ (grey/dotted curve). Both curves are almost
indistinguishable. In (a) all curves consist of $1000$ points, in (b) the
curves in the upper three panels as well as the grey curve in the fourth
consist of $600$ points each. $Q(b)$ in the fourth panel is different from the
one in the third panel with respect to having chosen $3000$ points.}
\label{fig6}
\end{figure}

Let us first remind of the fundamental property of {\em monotonicity} of any
dimension such as box-counting, which says that if $E$ is a subset of $F$ then
$dim(E)\le dim(F)$ \cite{Falc90,Tric95}. This property poses a strict upper
bound to any local dimension, which is the value of the dimension computed on
the respective larger interval that got subdivided. The numbers for the
box-counting dimension $B$ of the interval shown in Figs.\ \ref{fig1},
\ref{fig2} and for its blowups of Fig.\ \ref{fig3} appear to be consistent
with this inequality within the bounds of numerical errors. As explained in
Section IV.A, for our local $B$ we were not able to quantify the numerical
error, hence this consistency check indirectly yields some indication of the
size of these errors: In Figs.\ \ref{fig1} (b) and (c), \ref{fig3} (a) (which
was already noted to be problematic), and \ref{fig5} (a), (b) the local
fluctuations significantly exceed these upper bounds indicating that all
structures are mostly within the numerical errors. As was already discussed,
for the local autocorrelation dimension $A$ the numerical errors are even
larger. This is demonstrated by Fig.\ \ref{fig3} where the local $A$ of the
blowups partly drastically exceed the ones associated to Fig.\ \ref{fig2}
exemplifying again how delicate dimensional computations for these fractal
transport coefficients are, as we already remarked to the end of Section II.

In order to understand similarities and differences between fractal transport
coefficients and self-affine fractal functions we now elaborate on the
dimensionality of the Takagi function depicted in the inset of Fig.\
\ref{fig3} (b), which appears to be strikingly similar to the diffusion
coefficient shown in the same figure. First of all, as is proven in Ref.\
\cite{Falc90} for the Takagi function it is exactly $B=1$. However, this value
is identical with the respective lower bound for $B$, hence according to
monotonicity any subinterval must also have $B=1$. Consequently, the Takagi
function does not exhibit any irregularities in the local $B$. On the other
hand, the local difference quotient $Q$ of the Takagi function still
fluctuates irregularly due to the fact that this function is nowhere
differentiable. This appears to contradict the main result of Section IV.C
that there is an intimate relation between $B$ and $Q$ via $A$. However, this
relation is based, among others, on the assumption that $A$ decays like a
power law, see Eq.\ (\ref{eq:pld}), whereas numerical computations for the
Takagi function yield that $A$ decays exactly logarithmically. Consequently,
the autocorrelation dimension is not well-defined in this case and the link
between box-counting and the difference quotient breaks down.

At this point we may emphasize again that for our fractal transport
coefficients such as the Takagi-like one shown in Fig.\ \ref{fig3} (b) we
unambiguously find non-logarithmic decay for the autocorrelation function
corresponding to a power law for small enough variables. Further differences
between both graphs of Fig.\ \ref{fig3} (b) can be detected in the variations
of the local $Q$, which in case of the Takagi function are strictly
self-similar reflecting the construction of this function and which diverge
symmetrically to plus and minus infinity whenever this function shows a local
minimum. The local $Q$ of the diffusion coefficient, in comparison, is
inherently non-symmetric, as already indicated by the lower panel of Fig.\
\ref{fig3} (b), and diverges much more slowly to minus infinity than to plus
infinity. We also did preliminary computations for the local $B$ of the
Takagi function and compared them with our computations of the respective
diffusion coefficient. For the Takagi function there is an extremely poor
convergence to the box-counting power law Eq.\ (\ref{eq:bc}), and if by
mistake one computes local dimensions from these transient regimes one
erroneously generates fluctuations of the local $B$ that look analogous to the
fluctuations of the local $Q$. However, upon closer scrutiny one finds that
the different functions $N(\epsilon)$ assessing the local variation of $B$
according to the prescription in Section III.A slowly converge to each other
without any intersections. For our transport coefficient computations we did
not observe any such peculiar convergence of these functions in case of local
dimensions, instead we detected clear intersections indicating that we do not
obtain local fluctuations that are due to some initial transient regime. In
case of Fig.\ \ref{fig3} (b) these differences furthermore show up in the
quantitative values obtained for the local $B$, which are $1\le B\le 1.05$ for
the Takagi function, with larger portions of $B$ being close to one, and
approximately $1.02\le B \le 1.05$ for the diffusion coefficient.

We conclude that, although at first view the diffusion coefficient of Fig.\
\ref{fig3} (b) resembles very much a Takagi function, the latter has
significantly different dimensional properties: It is a fractal according to
our qualitative definition of a fractal in Section II, whereas to our
quantitative definition it provides a counterexample, since $B=1$.
Furthermore, the local $B$ is simply constant, $A$ is not well-defined and
hence, as far as we can tell, there is no simple functional relation between
the local $Q$ and any local dimension. It might be interesting to analyze
generalized Takagi functions with non-integer box-counting and autocorrelation
dimensions such as the ones introduced in Refs.\ \cite{Falc90,Tric95} along
the same lines. One may suspect that they yield further examples of
self-affine fractals that do not exhibit fractal fractal dimensions and for
which the simple relation to the local difference quotient suggested by Eq.\
(\ref{eq:CAQ2}) does not hold.

\section{Summary and conclusions}
The goal of this paper was to assess the irregular structure of the
parameter-dependent transport coefficients of a simple chaotic model system,
which is a two-parameter piecewise linear map defined on the line. For
computing these transport coefficients, which are the diffusion coefficient
and the current as functions of the slope and of the bias of the map, we used
a known numerical algorithm that is based on an exact solution of this problem
\cite{GrKl02}. The resulting data sets were evaluated by computing the box-counting
dimension and the dimension related to the autocorrelation function of the
respective graphs. Both quantities were computed {\em globally} for large
parameter intervals as well as {\em locally} on a regular grid of small but
finite subintervals yielding them as functions of the position of these
subintervals. Furthermore, we computed the local difference quotient on the
same set of subintervals.

Our main findings are that, firstly, both the diffusion coefficient and the
current of this model are fractal functions of the two control parameters, in
the quantitative sense that our numerical computations yielded box-counting
and autocorrelation dimensions for them that are, except in regions of phase
locking, larger than one. Secondly, we find that if both dimensions are
evaluated locally and plotted as functions of the position of the respective
subintervals they again exhibit nontrivial structure on finer and finer
scales, which is in agreement with our qualitative definition of fractality.
Computations of the box-counting dimension for the qualitatively similar local
difference quotient yielded values for these local variations that were
significantly larger than one. Hence, we concluded that both transport
coefficients are characterized by what we called {\em fractal fractal
dimensions}.  Thirdly, we detected a striking qualitative similarity between
the local box-counting dimension, the local autocorrelation dimension, and the
local difference quotient. While the relation between box-counting and the
autocorrelation function is standard, we showed by means of a simple
analytical approximation that there is an additional simple functional
relationship between the autocorrelation dimension and the difference
quotient. This result was verified numerically. A key ingredient was the
observation that not only these local quantities are fractal functions of the
control parameters but that furthermore another prefactor resulting from the
power law behaviour of the autocorrelation function turned out to be fractal
in the same way. This enabled to explain why partly the oscillations of the
local autocorrelation function and of the local box-counting dimension are
opposite to the ones of the absolute value of the difference quotient while in
other cases they are parallel to each other.

Finally, we applied the same analysis to the Takagi function in order to learn
to which extent the fractality of physical transport coefficients is similar
to, or different from, the one of self-affine fractal functions of generalized
Takagi, de Rham, or Weierstra{\ss} type. We found that the Takagi function
does not exhibit fractal fractal dimensions and that there is no functional
relationship to the local difference quotient, in sharp contrast to our
findings for the transport coefficients. As we discussed, it is extremely
difficult to obtain reliable numerical results if all dimensions are close to
one, however, our conclusion is that the fractal transport coefficients
analyzed in this paper belong to a different class of fractal functions than
the self-affine fractals of the type mentioned above. This is exemplified by
these significantly different properties and may be understood with respect to
the physical origin of the fractality of the transport coefficients. It would
be important to further study these similarities and differences by analyzing
other self-affine fractals along the same lines.

Another important, more mathematical problem concerns the limiting behaviour
of both local dimensions as well as of the local difference quotient in case
of fractal transport coefficients. It seems evident that, for all three
quantities, the respective limiting cases for the size of the subintervals
going to zero will not exist, however, it might be helpful to learn more about
the specific type of the supposed non-convergence possibly by looking at lower
and upper bounds for these functions in this limiting case. We furthermore
remark that similar irregular dependencies of local dimensions may hold not
only for the case of fractal functions but also for probability measures on
fractal sets such as attractors in dissipative systems. Another direction of
possible future research may concern the computation of power spectra for
these fractal transport coefficients in order to learn about the frequency
dependence of the fractal oscillations \cite{Falc90}. One may also think of
applying a wavelet analysis to these functions. We finally remark that
practically the local difference quotient is, of course, way easier to compute
than any local box-counting or autocorrelation dimension. Although at the
moment we see now straightforward way to obtain quantitative values for these
two local dimensions based on computing local difference quotients, since
there is no simple scaling between these different quantities, it appears that
the latter yields quite some information about the structure of fractal
transport coefficients. Hence, if one is primarily interested in the
qualitative fluctuations of local dimensions of the transport coefficients
studied in this paper the local difference quotient provides a convenient
access road to first approximate results.\\

\noindent {\bf Acknowledgments}\\
R.K.\ thanks J.Groeneveld, A.Pikovsky and M.V.Berry for inspiring remarks on
this subject that partly date back already some years ago. He is indebted to
K.Gelfert for very helpful hints on fractal dimensions that led to eliminating
a basic error in a previous manuscript. R.K.\ furthermore thanks Z.Koza for
interesting comments and for informing him about his parallel work on this
subject. He finally acknowledges a helpful hint by J.R.Dorfman and T.Gilbert
on Takagi functions and, together with T.K., thanks V.Reitmann for
discussions. The work by T.K.\ was performed during a three-month internship
at MPIPKS. Both R.K.\ and T.K.\ thank the MPIPKS for financial support of
T.K.\ during the last stages of this project.


\newpage

\vspace*{-1cm}

\end{document}